\documentclass[aip,pof,
 amsmath,
 amssymb,
]{revtex4-1}
\usepackage{graphicx}
\usepackage{dcolumn}% Align table columns on decimal point
\usepackage{bm}% bold math
\begin{document}
%\preprint{AIP/123-QED}
\title{Shock Waves in Dense Hard Disk Fluids}
\author{N. Sirmas}
\author{M. Tudorache}
\author{J. Barahona}
\author{M. I. Radulescu}
   \email{matei@uottawa.ca}
\affiliation{Department of Mechanical Engineering, University of Ottawa}
\date{\today}

\begin{abstract}
Media composed of colliding hard disks (2D) or hard spheres (3D) serve as good approximations for the collective hydrodynamic description of gases, liquids and granular media. In the present study, the compressible hydrodynamics and shock dynamics are studied for a two-dimensional hard-disk medium at both the continuum and discrete particle level descriptions. For the continuum description, closed form analytical expressions for the inviscid hydrodynamic description, shock Hugoniot, isentropic exponent and shock jump conditions were obtained using the Helfand equation of state. The closed-form analytical solutions permitted us to gain physical insight on the role of the material's density on its compressibility, i.e. how the medium compresses under mechanical loadings and sustains wave motion. Furthermore, the predictions were found in excellent agreement with calculations using the Event Driven Molecular Dynamic method involving 30,000 particles over the entire range of compressibility spanning the dilute ideal gas and liquid phases. In all cases, it was found that the energy imparted by the piston motion to the thermalized medium behind the propagating shock was quasi-independent of the medium's packing fraction, with a correction vanishing with increasing shock Mach numbers.
\end{abstract}

\maketitle
\section{Introduction}
\label{intro}
Dilute media composed of colliding hard disks (2D) and hard spheres (3D) serve as very good molecular models for gases, liquids and granular media.  Indeed, the kinetic theory of dilute hard sphere gases is well established \cite{Chapman&Cowling1970,Hirschfelder_etal1964}, whereby a hydrodynamic or coarse-grained description, such as the Navier-Stokes equations, can be obtained from the collective dynamics of many particles.  When the density of the hard particle media is increased, such that the distance travelled by a particle before experiencing another collision (i.e., the mean free path) becomes comparable with the dimension of the particle itself, departures from the ideal gas behavior become important.  For such regimes, hard particle media more closely resemble liquids and even solids. Significant work has thus been devoted to determine the equation of state of dense hard particle media \cite{Mulero_etal2008,Erpenbeck&Luban1985}, phase transition phenomena \cite{Alder&Wainright1957,Alder&Wainright1962,Zhao_etal2008} and shock wave structure \cite{Niki&Ono1977,Woo&Greber1999,Montanero_etal1998}.  With the advent of more powerful computers, hydrodynamic problems, such as the shock wave implosion,\cite{Gaspard&Lutsko2004} have been studied using hard sphere models.  Likewise, the hard particle description also permits to study complex problems in chemical kinetics and reactive dynamics \cite{Chou&Yip1984,Gorecki&Gorecka2000,Henriksen&Hanson2008} including detonation wave propagation \cite{Kawakatsu_etal1988,Erpenbeck1992}.  In all cases, the hard particle paradigm significantly simplifies the collision dynamics, such that kinetic-theory and statistical mechanics tools are more easily implemented to pass from the molecular (particle) dynamic description to the continuum description at larger scales.

One of the recent successes of the hard particle dynamics paradigm is to demonstrate how large hydrodynamic fluctuations can arise in granular media if one takes the collisions to be endothermic (dissipative)\cite{Goldhirsch&Zanetti1993}. Indeed, large scale clustering, spontaneous vortical flow formation, super-diffusion and departures from Maxwell-Boltzmann distributions arise when the collisions among particles are inelastic. These findings have triggered a renewed interest in the kinetic theory of hard particle media; a significant literature is now available on the subject, recently reviewed in Refs. \cite{Brilliantov&Poschel2004} and \cite{Goldhirsch2003}. The framework, however, is usually based on the dilute limit, although granular flows approach the structure of a liquid, or even solid, for which the compressibility is likely to severely affect the predictions. The present study addresses the dense regime of hard particle hydrodynamics.

Granular flows of macroscopic particles are typically highly compressible, in that the granular sound speed, which is approximately the mean speed of the particles, is usually comparable or smaller than the velocity of the solid surfaces driving the system.  For example, consider the walls of a vibrated container \cite{Swinney&Rericha2004,Goldshtein_etal1995,Ward_etal2005}, whose walls impart their energy to the colliding particles, or the propeller blades of a mechanical mill agitating grinding balls \cite{Reichardt&Wiechert2007}.  Likewise, the chute and avalanches of granular particles over obstacles \cite{Gray&Cui2007,Rericha_etal2002} may also generate strong shock waves, while the granular temperature and sound speed of the particles is very low owing to the low thermal velocity of the particles.  Such flows are thus characterized by the propagation of strong shock waves, propagating at large Mach numbers, often in the hypersonic regime.  The flow regime of granular media is thus highly compressible.

The present study focuses on the compressible dynamics of dense hard particle media, i.e. media in which the ideal gas assumption fails. We focus on the description of inviscid flows in such media, with a particular emphasis on shock wave solutions. Shock wave propagation, as generated by the sudden motion of a piston into a thermalized medium, is directly analogous to the sudden acceleration of a wall in a granular system \cite{Swinney&Rericha2004} or to the classical problem of a shock generated by a moving piston in gas dynamics \cite{Zeldovich&Raizer1966}.  We focus on large departures from the dilute limit and wish to describe the energy addition to the system by the piston's motion. Because we wish to compare the analytical results with molecular dynamic simulations and wish to easily visualize the system's dynamics, the present work focuses on the hard-disk system in two-dimensions.  This also permits to conduct calculations on much larger scales.  The restriction of our study to two space dimensions does not however affect the generality of the method and results, as they can be simply extended to three dimensions with the appropriate equation of state.  The present study also assumes the collisions to be elastic.  The shock propagation problem through a dissipative medium, for which continuum descriptions have been proposed in Ref. \cite{Kamenetsky_etal2000}, is left for future study.  Instead, we focus first on developing and validating the use of a sufficiently simple and accurate equation of state given by Helfand et al.\cite{Helfand&Frisch1961}, for which analytical solutions can be obtained for the dense regime offering a clear insight into the role of material's density on compressibility.  This will permit future extension to the dissipative regime.

The paper has the following structure.  In the first part, we use the simple equation of state for a hard disk fluid proposed by Helfand and Frisch\cite{Helfand&Frisch1961} to develop analytical descriptions for the medium's inviscid hydrodynamics and shock wave solutions.  In the second part of the paper, we report the results of molecular dynamic simulations and compare with the analytical predictions, hence further validating the model derived and further clarifying the role of the medium's density on its compressibility and shock dynamics.
%%%%%%%%%%%%%%%%%%   part 2   %%
\section{Continuum description}
\subsection{Equation of state for a hard disk gas}
\begin{figure}\includegraphics[width=0.5\textwidth]{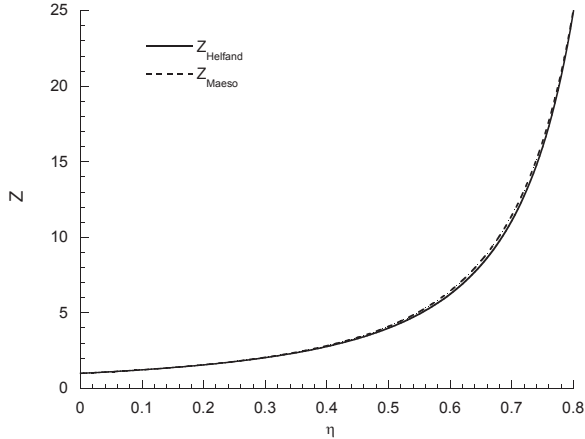}
  \caption{The relationship of the compressibility factor, \textit{Z}, with the packing factor $\eta$}
  \label{fig:EOS}
\end{figure}

In a hard particle system, the internal energy \textit{e} of the system consists only of the translational kinetic energy modes, with ${\frac{1}{2}}{\frac{k}{m}}{T}$ per translational degree of freedom, where \textit{k} is the Boltzmann constant, \textit{m} the particle mass, and \textit{T} the temperature.  For a hard disk system, there are only two translational degrees of freedom, yielding the caloric equation of state given by
\begin{equation}\label{e1}
e=2\left(\frac{1}{2}\frac{k}{m}T\right)
\end{equation}
An equation of state relation for the thermodynamic state variables $(p,v,T)$ also needs to be prescribed; generally, it can be written as \cite{Mulero_etal2008}:
\begin{equation}\label{e2}
pv=Z\frac{k}{m}T
\end{equation}
where $Z$ is the compressibility factor, which is unity for an ideal (dilute) gas.  Combining \eqref{e1} and \eqref{e2}, the internal energy can be written uniquely in terms of $p$ and $v$.
\begin{equation}\label{e3}  
e=\frac{pv}{Z}
\end{equation}
The compressibility factor $Z$ is usually expressed in terms of the packing factor $\eta$, denoting the fraction of the volume occupied by the particles, i.e.
\begin{equation}\label{eta}
\eta=\frac{V_a/m}{v}
\end{equation} 
where $V_a$ is the volume (i.e. the surface area of the disk in 2D) of a particle.  The caloric equation of state is thus uniquely expressed in terms of the pressure and specific volume in the form
\begin{equation}\label{e4}
e(p,v)=\frac{pv}{Z(\eta(v))}
\end{equation}
For a hard-disk medium, an exact equation of state is not known in closed form, hence requiring an infinite virial expansion. A thorough discussion of the various equations of state proposed for a hard disk system and their merits are described in Ref. \cite{Mulero_etal2008}. An equation of state that is both accurate, physically meaningful and analytically simple is that proposed by Helfand et al.\cite{Helfand&Frisch1961}.  It takes the form  
\begin{equation}\label{Helfand}
Z(\eta)=\frac{1}{(1-\eta)^2}
\end{equation}
To highlight its accuracy, we have compared it with the more accurate equation of state proposed by Maeso et al. \cite{Maeso1991}, which takes the form
%\begin{equation}\label{MSAV}
\begin{align}
Z(\eta)  = &  \frac{10^3-947.989\eta+128.018\eta^2- 113.987\eta^3}{(1-0.947989\eta)(1-\eta)^2 10^3} \nonumber\\
& +\frac{-52.9722\eta^4-1.580596\eta^5}{(1-0.947989\eta)(1-\eta)^2 10^3} \label{MSAV}
\end{align}
%\end{equation}
This equation of state is obtained from the first seven virial coefficients and has an accuracy characterized by an average absolute deviation of $0.05\%$.  The two equations of state are compared in Fig. \ref{fig:EOS}.
As can be verified from Fig. \ref{fig:EOS}, Helfand's equation of state offers a very good compromise between simplicity and accuracy over the entire range of compaction factors $\eta$.  It is generally accurate to within $\sim 3\%$ across the entire range of packing fractions spanning the dilute gas, dense gas, liquid and solid states close to the crystal packing fraction corresponding to a regular triangular lattice, given by
\begin{equation}\label{etac}
\eta_c=\frac{\pi \sqrt{3}}{6}\simeq 0.907
\end{equation}

We will henceforth adopt this equation of state for analytical simplicity and insight into the effect of initial density (packing factor) on the medium's compressibility.
%%%%%%%%%%%%%%%%%%%%%%%%%%%%%%%%%%%%%%%%%%%%%%%%%%%%%%%%%%
\subsection{Isentropic exponent and sound speed}
The change in the compressibility of the medium is best described by the isentropic exponent $\gamma$, which describes the relation between changes in density and changes in pressure for isentropic processes. It is defined as
\begin{align}
\gamma \equiv \left( \frac{\partial lnp}{\partial ln\rho} \right)_s \label{eq:gamma}
\end{align}
where $\rho=1/v$ is the density.  The sound speed in the medium is directly related to $\gamma$, i.e.:
\begin{align}
c^2 \equiv \left( \frac{\partial p}{\partial \rho} \right)_s =\gamma \frac{p}{\rho} \label{csqr1}
\end{align}

\begin{figure}\includegraphics[width=0.5\textwidth]{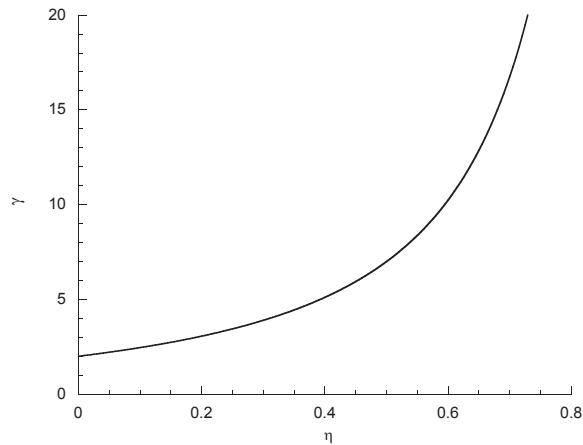}
  \caption{The variation of the isentropic exponent with the local packing factor $\eta$}
  \label{fig:gamma}
\end{figure}

Note that $\gamma$ is \textit{not} the ratio of specific heats, which in this case is simply 2 for any compaction ratio, as can be deduced from \eqref{e1} and the definition of enthalpy.  To determine the dependence of $c$ or $\gamma$ on the state variables $p$ and $v$ (or $\rho$), we have to make use of the Gibbs thermodynamic equations relating $p$, $v$ and the entropy $s$.  The first and second law of thermodynamics for a control system yield Gibbs equation written in terms of the density as
\begin{equation}\label{Gibbs}
de(p,\rho)=Tds+\frac{p}{\rho^2}d\rho 
\end{equation}
alternatively, $e(p,\rho)$ can also be written as a perfect differential as
\begin{equation}\label{Gibbs2}
de(p,\rho)={\left( \frac{\partial e}{\partial p} \right) }_{\rho}dp+{\left( \frac{\partial e}{\partial \rho} \right)}_{p}d\rho
\end{equation}
Comparing \eqref{Gibbs} and \eqref{Gibbs2}, we can re-write
\begin{equation}\label{Gibbs3}
dp= \frac{ \frac{p}{{\rho}^2} -{\left( \frac{\partial e}{\partial \rho} \right)}_{p}}{{\left( \frac{\partial e}{\partial p} \right)}_{\rho}} d\rho+
\frac{T}{{\left( \frac{\partial e}{\partial p} \right) }_{\rho}}ds
\end{equation}
Now writing $p(\rho, s)$ as a perfect differential, we get
\begin{equation}\label{prhos}
dp(\rho, s)={\left( \frac{\partial p}{\partial \rho} \right)}_{s}d\rho+{\left( \frac{\partial p}{\partial s} \right)}_{\rho}ds
\end{equation}
Comparing \eqref{Gibbs3} with \eqref{prhos}, we get immediately that
\begin{equation}\label{sound1}
c^2={\left( \frac{\partial p}{\partial \rho}\right)}_{s}=\frac{\frac{p}{{\rho}^2} -{(\frac{\partial e}{\partial \rho})}_{p}}{{\left( \frac{\partial e}{\partial p} \right)}_{\rho}}
\end{equation}
This expression is general for any medium, as we have not yet used the equation of state of the medium.  Using expressions \eqref{eta}, \eqref{e4} and \eqref{Helfand} to evaluate the derivatives, we get immediately the expression for the sound speed in terms of the local state of the medium and the local packing factor $\eta$
\begin{equation}\label{sound}
c^2=pv \left( 1+(1-\eta)^{-2}+2\eta(1-\eta)^{-1} \right)
\end{equation}
and from \eqref{csqr1} we get the isentropic exponent.
\begin{equation}\label{gamma2}
\gamma=1+(1-\eta)^{-2}+2\eta(1-\eta)^{-1}
\end{equation}
Figure \ref{fig:gamma} shows the variation of the isentropic exponent with the local packing fraction.  In the dilute limit of an ideal gas, i.e. $ \eta \rightarrow 0 $, we recover the isentropic exponent of a 2-dimensional hard disk gas of $\gamma = 2$.  However, with increasing packing fraction, the isentropic exponent grows commensurably, reflecting the incompressibility of the medium.  A small change in density requires a very large change in pressure, a characteristic of nearly \textit{incompressible} media like liquids and solids.

We stress out that having written the isentropic exponent in closed form permits us to formulate the hydrodynamic equations for the medium very simply.  This is performed in the next section.
%%%%%%%%%%%%%%%%%%%%%%%%%%%%%%%%
\subsection{Gasdynamic description of a hard disk medium}

The compressible motion of a medium, after neglecting slow processes involving molecular transport, are governed by the Euler equations.  Since the resulting motion is particle isentropic, we can write them as:
\begin{subequations}
\begin{align}
&\frac{D\rho}{Dt} = -\rho\nabla\cdot\textbf{u}\\
&\rho \frac{D\textbf{u}}{Dt} = -\nabla p\\
&\frac{Ds}{Dt} =0 \label{eq:partisen}
\end{align}
\end{subequations}
where $\textbf{u}$ is the macroscopic velocity vector and $D/Dt=\partial_t +\textbf{u} \cdot \nabla$ is the usual material derivative, following an element of fluid.  Using \eqref{prhos}, applying it on the particle path (i.e., following a thermodynamic system) and making use of \eqref{csqr1} and \eqref{eq:gamma}, one can re-express \eqref{eq:partisen} as 
\begin{equation}
\frac{D ln (p)}{Dt}=\gamma(\rho)\frac{D ln (\rho)}{Dt}
\end{equation}
Using the relation for $\gamma$ given by \eqref{gamma2}, this can be integrated in closed form in order to determine the isentropic relation between pressure and density along a particle path.  Taking the reference state as $p_1$ and $\rho_1$, the isentrope becomes
\begin{align}
\frac{p}{p_1}={\left(\frac{\rho}{\rho_1}\right)}^2 \left( \frac{1-\frac{\rho}{\rho_1}\eta_1}{1-\eta_1} \right)^{-3} e^\frac{\left(\frac{\rho}{\rho_1}-1\right)\eta_1}{\left(1-\frac{\rho}{\rho_1}\eta_1 \right) \left(1-\eta_1 \right)} 
\end{align}
where we recover the result obtained by Gaspard and Lutsko\cite{Gaspard&Lutsko2004} using a more complex equation of state.  For ideal hard-disk gases, in the limit of vanishing packing factor, we recover the isentropic relation $p \propto \rho^2$.

\subsection{Shock Waves in a Hard-Disk Medium}
\begin{figure}
  \includegraphics[width=0.3\textwidth]{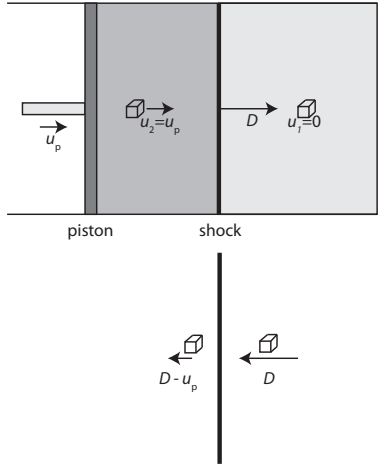}
\caption{Propagation of a piston driven shock wave into a quiescent medium and transformation to the shock fixed frame of reference (below)}
\label{fig:setup}
\end{figure}

When shock waves are generated in a medium, weak solutions to the inviscid Euler equations are required.  Consider a piston suddenly accelerated from zero velocity to a constant velocity $u=u_p$ into a medium initially at rest; see Fig. \ref{fig:setup}.  A shock wave is formed, which propagates with velocity $D$.  If diffusive fluxes are negligible in the pre- and post-shock states, and letting the subscript 2 denote the uniform state of the medium behind the shock wave, the conservation of mass, linear momentum and energy, in the frame of reference of the shock (see Fig. \ref{fig:setup}) yield\cite{Zeldovich&Raizer1966}:
\begin{align}
\frac{D}{v_1} = \frac{(D - u_p)}{v_2}\label{mass} \\ 
p_1+\frac{D^2}{v_1} = p_2+\frac{(D - u_p)^2}{v_2} \label{momentum}\\ 
e_1+p_1 v_1+\frac{1}{2} D^2 = e_2+p_2 v_2+\frac{1}{2} (D - u_p)^2 \label{energy}
\end{align} 
where $v$ is the specific volume, $p$ is the hydrodynamic pressure and $e$ is the specific internal energy.
%%%%%%%Fig%%%%%%%%%%%
%%%%%%%%%%%%%%%%%%%%%
Combining the equation of mass \eqref{mass}, momentum \eqref{momentum} and energy \eqref{energy} to eliminate the two speeds, the accessible end states 2 across the shock wave can be represented by the shock Hugoniot relation in the $(p,v)$ plane\cite{Zeldovich&Raizer1966}:
\begin{equation}\label{Hugo}
e_2(p_2, v_2)-e_1(p_1, v_1)=\frac{1}{2} (v_1-v_2 )(p_1+p_2)
\end{equation} 
with either $D$ or $u_p$ as parameters from \eqref{mass} and \eqref{momentum}. So far, the shock description is completely general, and depends on the equation of state $e(p,v)$ characterizing the particular medium.
%%%%%%%%%%%%%%%%%%%%%%%%%%%%

\begin{figure}\includegraphics[width=0.5\textwidth]{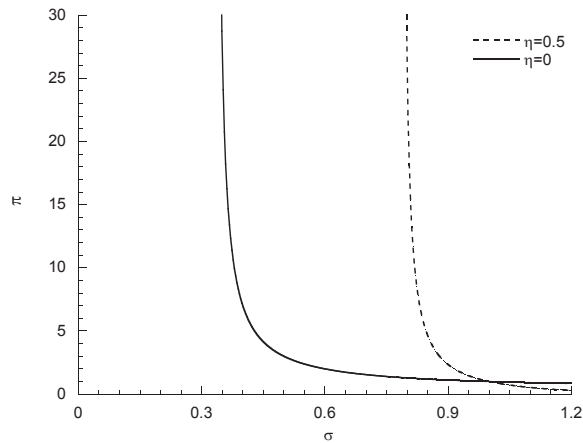}
  \caption{The relation between the shock overpressure and compression ratio represented on the shock Hugoniot for a hard disk medium in terms of the initial packing factor $\eta_1$}
  \label{fig:hugoniot}
\end{figure}

\begin{figure}\includegraphics[width=0.5\textwidth]{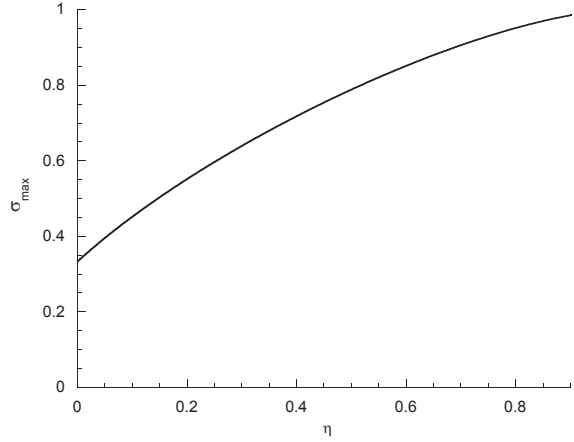}
  \caption{The maximum compression ratio achieved by shock compression in terms of the initial packing factor $\eta_1$}
  \label{fig:maxcompression}
\end{figure}

Using Helfand's equation of state \eqref{Helfand} and auxiliary relations \eqref{eta} and \eqref{e4}, the Hugoniot relation \eqref{Hugo} can be used to express in closed form the variation of pressure with specific volume across the shock wave, yielding
\begin{equation}\label{PiHugo}
\pi=\frac{\frac{1}{2} (1-\sigma)+(1-\eta_1)^2}{\sigma(1-\frac{\eta_1}{\sigma})^2-\frac{1}{2} (1-\sigma)}
\end{equation}
where $\pi$, $\sigma$ and $\eta_1$ are, respectively, the non-dimensional pressure ratio across the shock wave, non-dimensional specific volume ratio across the shock wave, and the initial packing fraction of the medium.
\begin{equation}\label{nonD}
\pi=\frac{p_2}{p_1}, \sigma= \frac{v_2}{v_1}, \eta_1=\frac{(V_a/m)}{v_1}
\end{equation}
Figure \ref{fig:hugoniot} displays the shock Hugoniot curves for two initial packing fractions.  As can be seen, as the packing fraction is increased, a higher pressure is generated for the same compression ratio.  Alternatively, the same pressure is achieved with less compression in an initially higher packed medium. This mimics very well the real properties of shocks in less compressible media such as liquids and solids \cite{Zeldovich&Raizer1966}.
The maximum compression that can be achieved across a shock wave can be found by letting $ \pi \rightarrow \infty $ in \eqref{PiHugo}.  This corresponds to letting the denominator in \eqref{PiHugo} vanish, yielding a quadratic for which the largest root is the physically attainable solution.  The maximum compression achievable is thus
\begin{equation}\label{sigmamax}
\sigma_{max}=\frac{1}{6} (1+4 \eta_1 + \sqrt{(1-8( \eta_1 - 1) \eta_1})
\end{equation}
which is shown graphically in Figure \ref{fig:maxcompression}.
%%%%%%%%%%%%%FIG%%%%%%%%%%%%%%%
%%%%%%%%%%%%%%%%%%%%%%%%%%%%%%%%%%%%%%%%%%%%%%%%%%%%%%%%%%%%

\begin{figure}\includegraphics[width=0.5\textwidth]{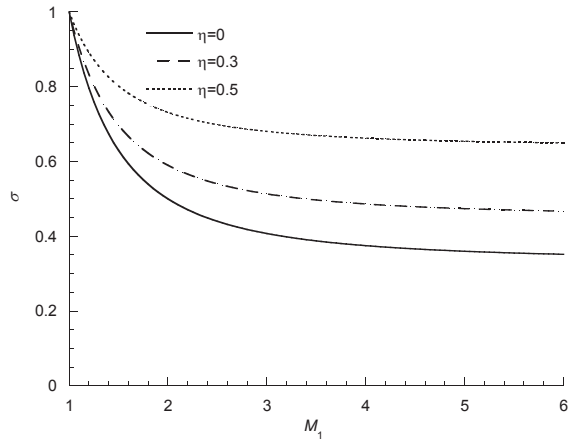}
  \caption{The variation of the compression ratio with the shock Mach number for different initial packing factors}
  \label{fig:sigma}
\end{figure}

\begin{figure}\includegraphics[width=0.5\textwidth]{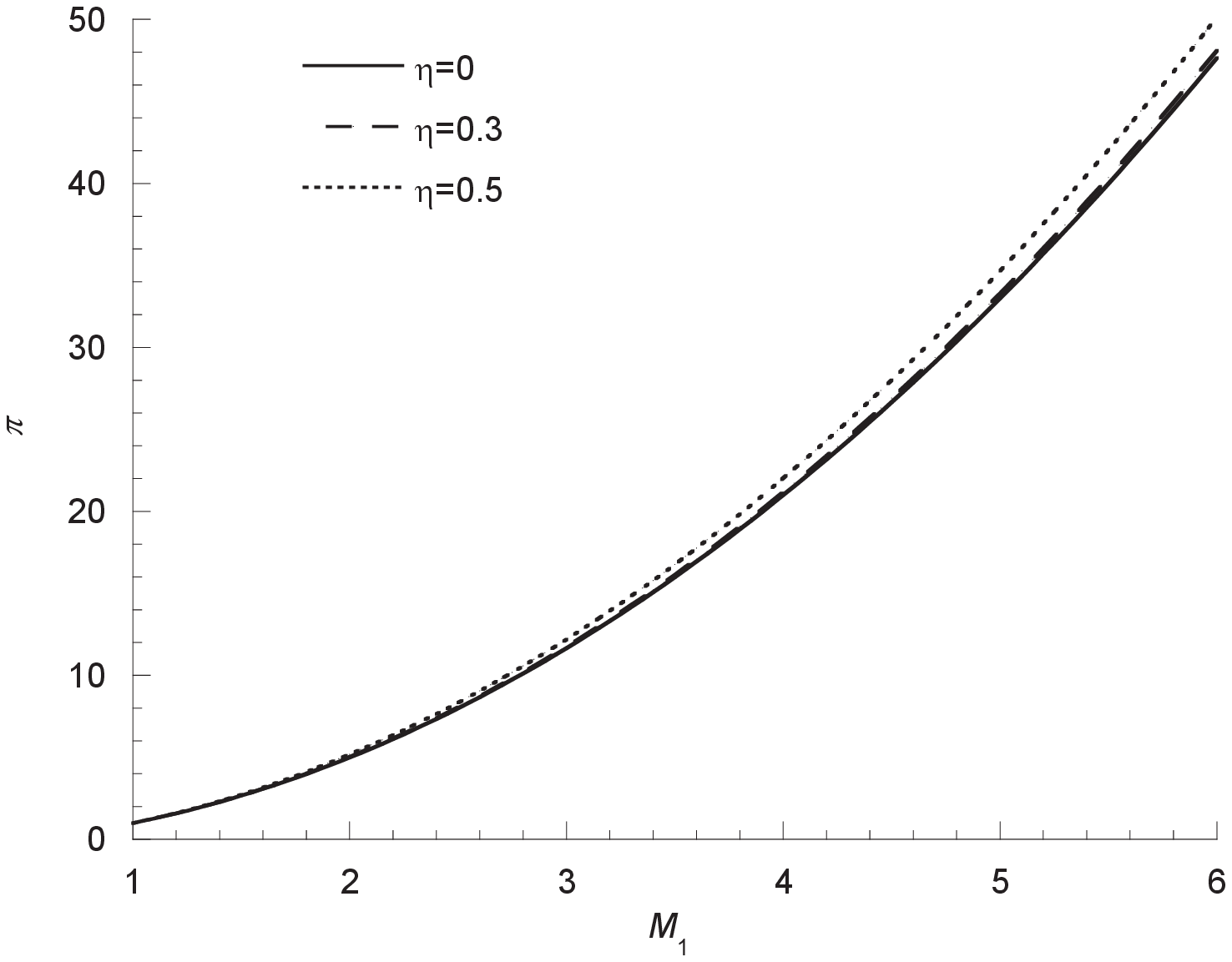}
  \caption{The variation of the pressure ratio with the shock Mach number for different initial packing factors}
  \label{fig:pi}
\end{figure}

\begin{figure}\includegraphics[width=0.5\textwidth]{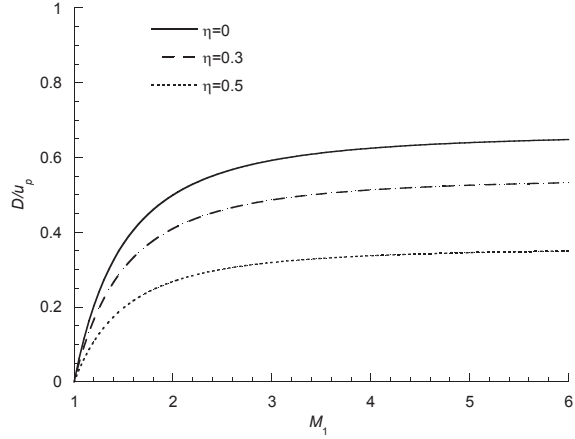}
  \caption{The variation of the piston to shock speed ratio with the shock Mach number for different initial packing factors}
  \label{fig:up}
\end{figure}

\begin{figure}\includegraphics[width=0.5\textwidth]{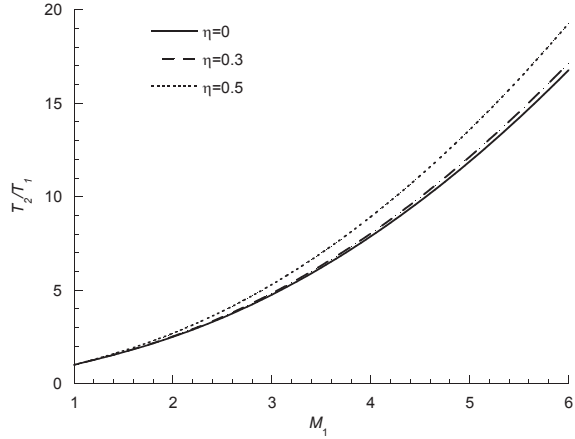}
  \caption{The variation of the temperature ratio with the shock Mach number for different initial packing factors}
  \label{fig:temp}
\end{figure}

It is useful to parametrize the shock jump conditions based on the Mach number of the shock wave $M_1$.  From the mass \eqref{mass} and momentum \eqref{momentum} equations and using \eqref{eta}, \eqref{e4}, \eqref{Helfand} and \eqref{nonD}, we can express the overpressure across the shock wave in terms of the shock Mach number and compression ratio across the shock, yielding the so-called Rayleigh line
\begin{equation}\label{Rayleigh}
\pi=1+\psi(1-\sigma)
\end{equation}
where $\psi$ is simply related to the shock Mach number and the initial value of $\gamma$:
\begin{equation}\label{psi}
\psi\equiv\frac{D^2}{{p}_1{v}_1}={\gamma_1} \frac{D^2}{{c_1}^2}={\gamma_1} {{M_1}^2}
\end{equation}
Equating \eqref{PiHugo} to \eqref{Rayleigh}, we obtain the shock jump conditions in closed form in terms of the shock strength parameter $\psi$, and the initial packing fraction $\eta_1$:
\begin{subequations}\label{jumps}
\begin{align}
\frac{v_2}{v_1} & =  \sigma= \frac{4+\psi+4\psi+4\psi\eta_1}{6\psi} \nonumber\\
 & +\frac{\sqrt{-24\psi(1+\psi)\eta_1^2+(4+\psi+4\psi\eta_1)^2}}{6\psi}\\
\frac{p_2}{p_1} & = \pi = 1+\psi(1-\sigma)\\
\frac{u_p}{D} & =  1-\sigma \label{upD}\\
\frac{T_2}{T_1} & =  \pi\sigma \frac{{1-\frac{\eta_1}{\sigma}}^2}{(1-\eta_1)^2}
\end{align}
\end{subequations}
Figures \ref{fig:sigma}, \ref{fig:pi}, \ref{fig:up}, \ref{fig:temp} show these relations graphically for $\eta_1=0$, $\eta_1=0.3$ and $\eta_1=0.5$ in terms of the shock Mach number $M_1$.  For a given compression ratio $\sigma$, increasing the initial compaction $\eta_1$ gives rise to larger shock Mach numbers.  The temperature and pressure ratios, however, are not very sensitive to changes in the initial compaction levels, as can be observed from Figures \ref{fig:pi} and \ref{fig:temp}.
%%%%%%%%%%%%%%%%%%%%%%%%%%%%%%%%%%%%%%%%%%%%%% 

The parametrization of the shock jump conditions by the shock Mach number, although practical, is however not very insightful into the physics of the problem, which correspond to the response of a medium to a given prescribed compression.  Instead, it is useful to choose the square of the piston speed, representing the kinetic energy addition by the piston motion as a parameter, given by  
\begin{equation}\label{beta}
\beta\equiv \frac{{u_p}^2}{e_1} 
\end{equation}
Expressions \eqref{jumps} can be expressed implicitly in terms of $\beta$ using \eqref{upD}, which can be rewritten as
\begin{equation}
\beta=\psi \frac{(1-\sigma)^2}{(1-\eta_1)^2}
\end{equation}
%%%%%%%%%%%%%%%%%%%%%%%%%%%%%%%%
\begin{figure}\includegraphics[width=0.5\textwidth]{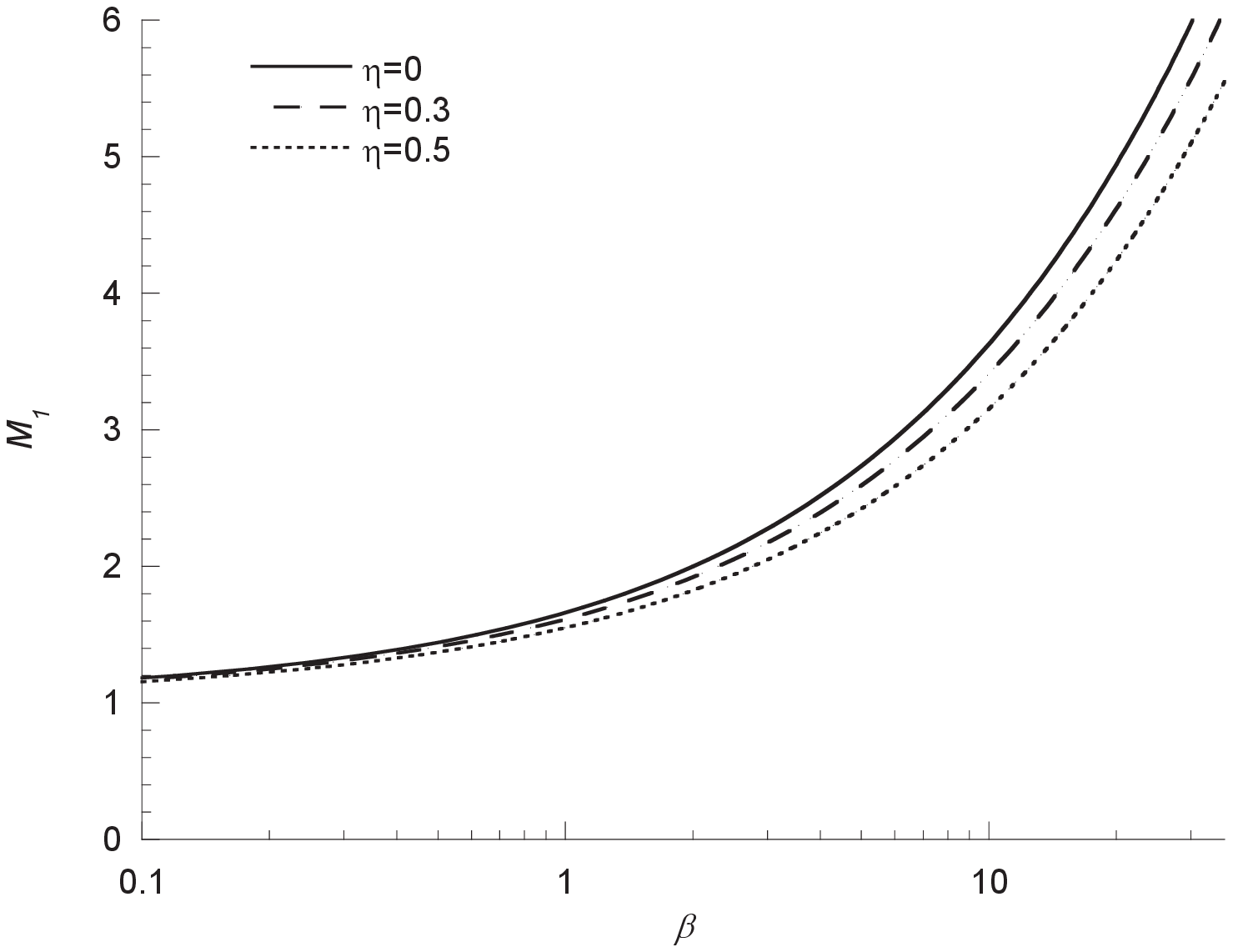}
  \caption{The variation of the shock Mach number with $\beta$ for different initial packing factors}
  \label{fig:MachU}
\end{figure}
\begin{figure}\includegraphics[width=0.5\textwidth]{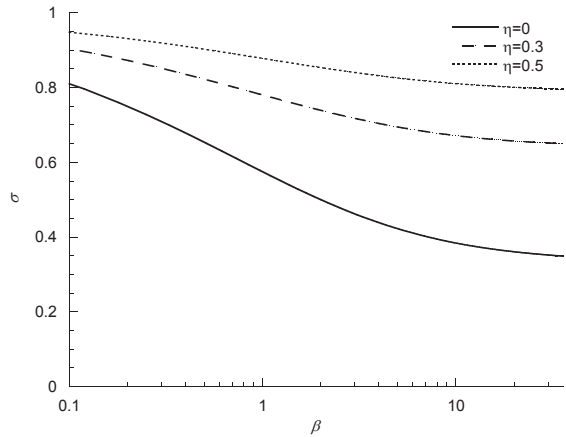}
  \caption{The variation of the compression ratio with $\beta$ for different initial packing factors}
  \label{fig:sigmaU}
\end{figure}
\begin{figure}\includegraphics[width=0.5\textwidth]{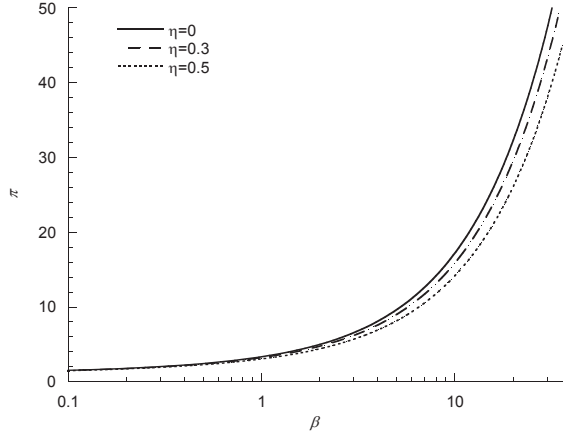}
  \caption{The variation of the shock overpressure with $\beta$ for different initial packing factors}
  \label{fig:piU}
\end{figure}
\begin{figure}\includegraphics[width=0.5\textwidth]{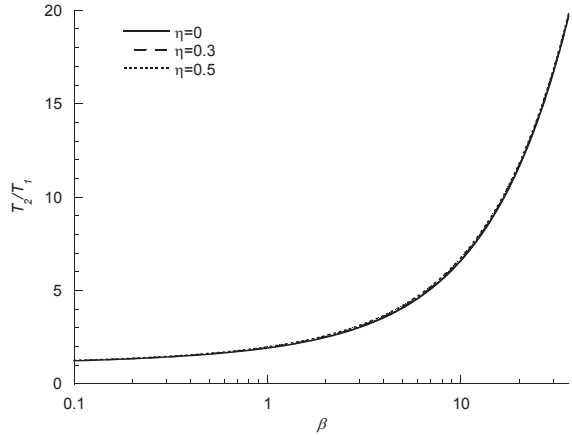}
  \caption{The variation of the temperature ratio with $\beta$ for different initial packing factors}
  \label{fig:tempU}
\end{figure}
Figures \ref{fig:MachU}, \ref{fig:sigmaU}, \ref{fig:piU}, \ref{fig:tempU} show the variation of the shock Mach number, specific volume ratio, pressure ratio and temperature ratio with $\beta$ varying over more than two orders of magnitude (shown in logarithmic scale for clarity).  For the same piston speed, increasing the initial compaction ratio and hence departing from the dilute gas approximation, a weaker shock is transmitted with a lower pressure ratio, smaller volumetric compression of the medium and lower Mach number.  Surprisingly, however, the temperature ratio shown in Fig. \ref{fig:tempU} for a given piston speed is only very weakly affected by the departure from the dilute gas across the entire range of weak to strong shocks.  Instead, even in the closely packed regimes, the temperature ratio is very well approximated by the ideal gas expression, even for the weak shocks. To account for this novel finding, the temperature ratio can be written directly from the Hugoniot relation \eqref{Hugo} which, after some manipulation, can be rewritten as
\begin{equation}\label{ejump}
\frac{T_2}{T_1}=\frac{e_2}{e_1}=1+\frac{1-\sigma}{(1-\eta_1)^2}+\frac{1}{2}\frac{{u_p}^2}{e_1}  
\end{equation}
The second term on the right hand side of \eqref{ejump}, labeled A, is bounded as $u_p \rightarrow \infty$, while the third term, labeled B, grows without bounds as $u_p \rightarrow \infty$, hence dominates the temperature increase across the shock wave.  Formally taking the ratio of these two terms, we obtain
\begin{equation}\label{condition}
\frac{A}{B}=\frac{2}{(1-\sigma)\gamma M^2} 
\end{equation}
In the limit of a large shock Mach number, $\sigma \rightarrow \sigma_{max}$ (from \eqref{sigmamax}) and hence $A/B \rightarrow 0$. Likewise, this term also vanishes for arbitrary shock Mach numbers when the initial compaction ratio is increased.  Indeed, the isentropic exponent increases very rapidly with packing factor, as shown in Fig. \ref{fig:gamma}.  Under these conditions, the internal energy and temperature, according to \eqref{ejump}, are thus incremented by the kinetic energy of the piston, resulting into an equipartion of the available energy into the mean motion (mean kinetic energy of the medium) and thermal energy.  
\begin{equation}\label{Tinv}
\frac{T_2}{T_1}=\frac{e_2}{e_1} \cong 1+\frac{1}{2}\frac{{u_p}^2}{e_1} 
\end{equation}
This result was also obtained in Ref. \cite{Zeldovich&Raizer1966} by arbitrarily neglecting the upstream pressure of an arbitrary medium in the governing equations. Instead, the condition that the term given in \eqref{condition} be small can be considered as the strong shock condition in a hard disk fluid. Generally, it was numerically found that the correction for low Mach numbers is very weak, as can be seen in Figure \ref{fig:tempU}, suggesting that \eqref{Tinv} serves as a very good approximation for all initial compaction ratios of the medium, ranging from the dilute ideal gas to a solid packed medium of hard particles.
%%%%%%%%%%%  Section 3   %%%%%%%%%%%%%%%%%%%%%%%%%% 
\section{Comparison with Molecular dynamic calculations}
In order to validate the results obtained analytically for the shocked state following the shock compression by a moving piston, we have performed a series of particle dynamic simulations.  Owing to the fact that hard particles interact during a vanishingly small time compared to their flight time, during which their trajectories are un-affected by any forces, their dynamics can be solved in closed form \cite{Alder&Wainright1959}. In between collisions, the particle trajectories are known exactly, while the infinitely short collisions only modify the velocity of particles upon collision.  The collision rules are such that the binary system of colliding particles conserves its linear momentum and energy.  The sequence of collisions can thus be handled quite efficiently in an algorithm evolving the system of many particles collision after collision with only very minimal loss in accuracy. This solution strategy is called the Event Driven Molecular Dynamics method (EDMD).

\begin{figure}\includegraphics[width=0.4\textwidth]{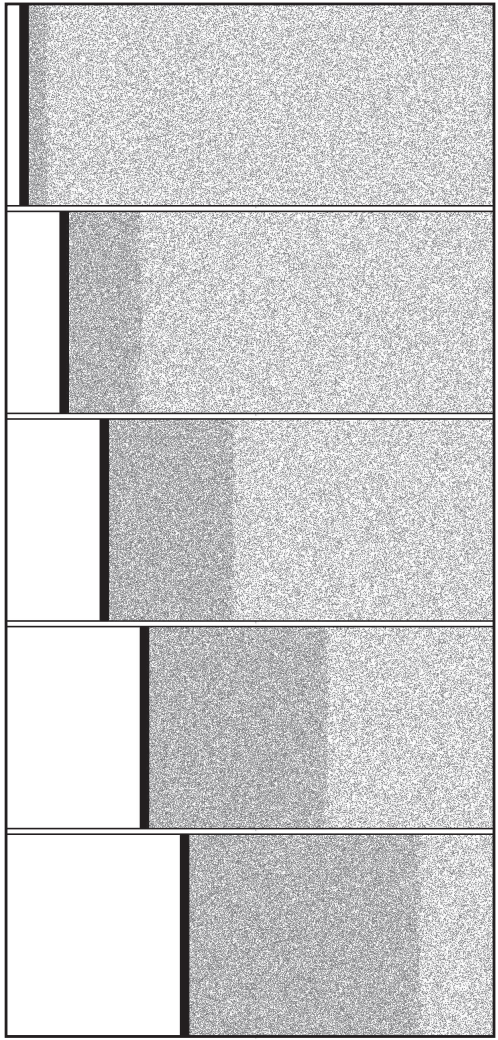}
  \caption{Five consecutive snapshots illustrating the shock wave driven by the moving piston for $\beta=6.25$ and $\eta_1=0.192$}
  \label{fig:shockMD}
\end{figure}

\begin{figure}\includegraphics[width=0.5\textwidth]{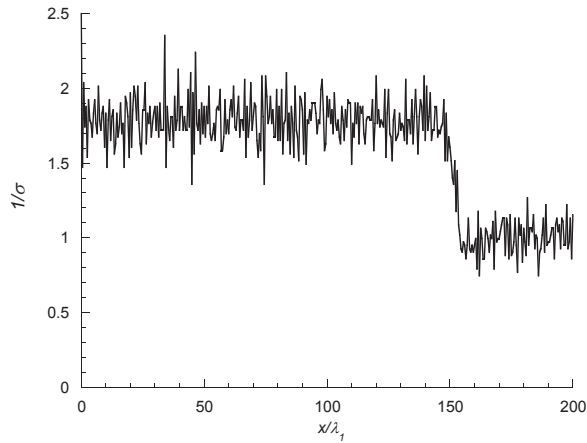}
  \caption{The density profile captured for a single realization for $\beta=16$ and $\eta_1=0.192$}
  \label{fig:densityMDsingle}
\end{figure}

\begin{figure}\includegraphics[width=0.5\textwidth]{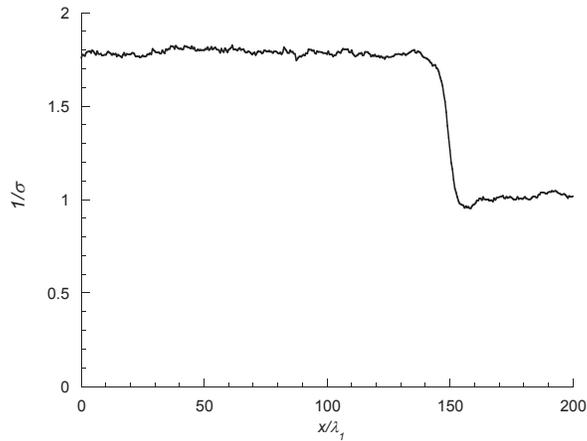}
  \caption{The average density profile obtained for $\beta=16$ and $\eta_1=0.192$}
  \label{fig:densityMD}
\end{figure}

\begin{figure}\includegraphics[width=0.5\textwidth]{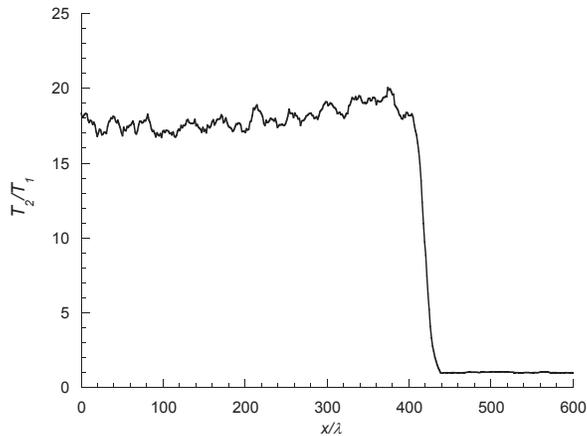}
  \caption{The average temperature profile obtained for $\beta=16$ and $\eta_1=0.192$}
  \label{fig:tempMD}
\end{figure}

The calculations performed in the present study use the implementation of the EDMD method of Poschel and Schwager for two dimensional hard-disk systems \cite{Poschel&Schwager2005}.  Consistent with the theoretical treatment above, the rotation of the particles is not considered.  The code developed by Poschel and Schwager was modified to allow for one moving wall, while retaining the event-driven nature of the algorithm. The calculations were performed in a rectangular domain of dimensions $L_x$ by $L_y$.  Particle reflections at all the walls were taken as elastic and specular. A total number of 30,000 particles were used.  Controlling the dimension of the domain set the initial specific volume of the system $v_1$.  Adjustment of the radius of the particles controlled the initial packing factor of the system $\eta_1$ given by  
\begin{align}
\eta_1 = \frac{N V_a}{V} = \frac{N \pi {d}^2}{4 L_x L_y}
\end{align}
The particles were first positioned randomly in the domain, ensuring no overlapping.  All the particles were assigned equal speeds, with random directions. This initial energy provided to the system corresponds to the initial specific energy of the system $e_1$.\\
\begin{equation}
e_1=\frac{1}{2} <U_1^2>
\end{equation}
which, according to \eqref{e1}, sets the initial temperature of the system. The system of particles was then let to equilibrate while keeping the number of particles, volume and energy constant (i.e., the NVE micro-canonical ensemble), until the distribution of speeds approached the Maxwell-Boltzmann distribution.  Once the system of particles reached thermal equilibrium, the left wall was set in motion with constant velocity $u_p$.  An example of the following dynamics is shown in Figure \ref{fig:shockMD}.  A clearly demarked hydrodynamic shock can be seen to propagate across the system of hard disks.
In the results reported below, all distances have been normalized by the diameter of the hard disks, masses by the atomic mass and velocities by the initial speed assigned to the particles $U_1$.  Table 1 gives the set of three initial conditions investigated, corresponding respectively to a dilute gas ($\eta_1=0.012$), dense gas ($\eta_1=0.192$) and much denser liquid state ($\eta_1=0.433$).  Also listed in Table 1 is the mean free path, estimated from the kinetic theory of a hard disk medium \cite{Mulero_etal2008} by
\begin{equation}\label{meanfreepath}
\lambda= \frac{L_x L_y}{2\sqrt{2} N d } 
\end{equation}
%%%%%%%%%%%%%%%TABLE 1%%%%%%%%%%%%%%%%%%%%%%%%%
\begin{table}
\vspace{-10pt}
\caption{Parameters for the three fluid regimes investigated via EDMD}
\label{tab:params}
\begin{tabular}{c c c}
\hline\noalign{\smallskip}
Domain size & Packing fraction & Mean free path\\ 
$L_x$ x $L_y$ & $\eta_1$ & $\lambda_1$\\
\noalign{\smallskip}\hline\noalign{\smallskip}
1960 x 1000 & 0.012 & 23.1\\
490 x 250 & 0.192 & 1.44\\
327 x 167 & 0.433 & 0.642\\
\noalign{\smallskip}\hline
\end{tabular}
\vspace{-10pt}
\end{table}
%%%%%%%%%%%%%%%%%%%%%%%%%%%%%%%%%%%%%%%%%%%%%%%%%%%
For each packing fraction, the speed of the piston was modified in order to achieve different levels of compression.  In all simulations, the density and temperature were obtained by coarse-grained averages of the state of the gas.  At one particular time of the simulation, the domain was separated into strips $0.5 \lambda_1$ wide.  In each strip, the density (or specific volume) was directly obtained by counting particles whose center of mass was within the strip.  The internal energy, hence temperature, was obtained by taking the average of the kinetic energies of each particle comprised in the strip and subtracting the mean kinetic energy in the macroscopic motion, i.e.; 
\begin{equation}
T=\frac{\frac{1}{2}\left< U-\left< U_x \right> \right> ^2}{k/m}
\end{equation}
Because the position of the shock is not known a priori, we have used ensemble averages to determine more accurate measurements of the coarse-grained density and temperature.  This was achieved by repeating the calculation with statistically different initial conditions, which were implemented by starting the piston motion at slightly different times after the initial equilibration of the medium.  An example of the density profile captured during a single realization is shown in Figure \ref{fig:densityMDsingle}. The  ensemble average of 5 of these realizations for density and temperature profiles are shown in Figures \ref{fig:densityMD} and \ref{fig:tempMD}. As can be seen, the density and temperature relax slowly to their equilibrium values behind the shock jump.  Note that the internal structure of the shock is not within the scope of the present study, which deals exclusively with the jump conditions across the shock wave in the thermally relaxed media. The results reported below provide the equilibrated values across the shock wave.  Using the density profiles, we have also determined the position of the shock wave at successive times by tracking the position where the density increases by $50\%$ of its total increase. This permitted to determine the shock speed.\\

%%%%%%%%%%%%%%%%%%%%%%%%%%%

%%%%%%%%%%%%%%%%%%%%%%%%%%%%

%%%%%%%%%%%%%%%%%%%%%%%%%%%%
\begin{figure}\includegraphics[width=0.5\textwidth]{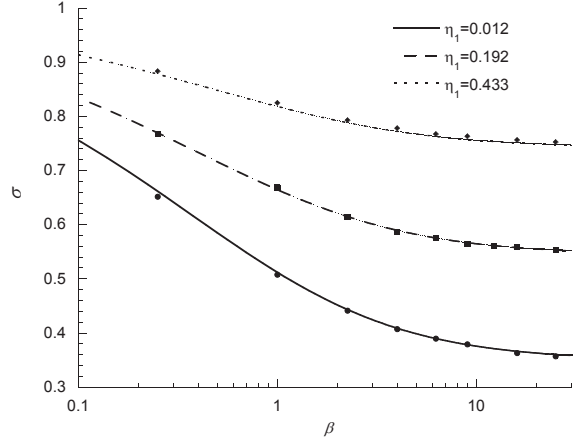}
  \caption{Comparison between analytical and calculated compression ratios for different initial packing factors and varying shock strength}
  \label{fig:sigmaCOMP}
\end{figure}

%%%%%%%%%%%%%%%%%%%%%%%%%%%%
\begin{figure}\includegraphics[width=0.5\textwidth]{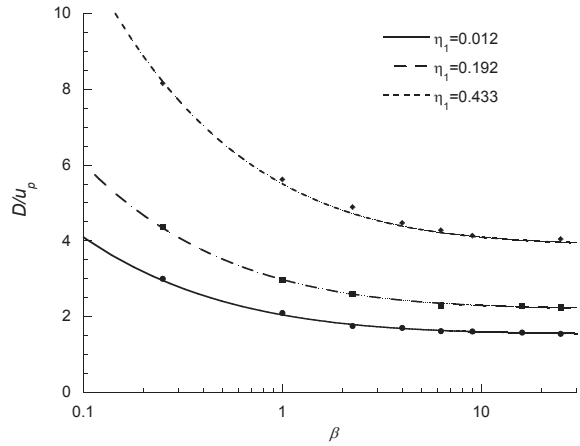}
  \caption{Comparison between analytical and calculated piston speeds for different initial packing factors and varying shock strength}
  \label{fig:uCOMP}
\end{figure}

%%%%%%%%%%%%%%%%%%%%%%%%%%%%
\begin{figure}\includegraphics[width=0.5\textwidth]{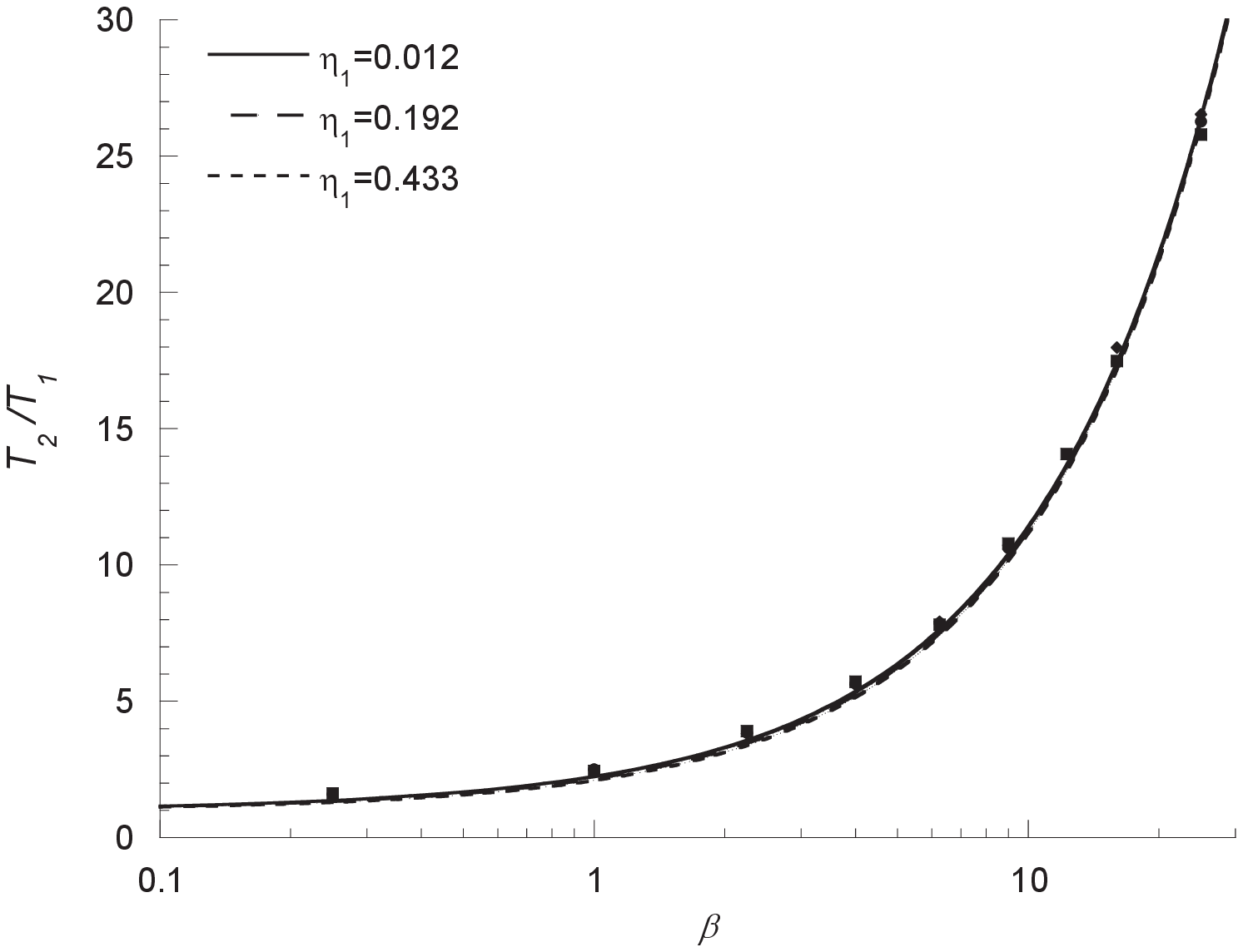}
  \caption{Comparison between analytical and calculated temperature ratios for different initial packing factors and varying shock strength; symbols same as Figure \ref{fig:uCOMP}}
  \label{fig:Tcomp}
\end{figure}
%%%%%%%%%%%%%%%%%%%%%%%%%%%%

Figures \ref{fig:sigmaCOMP}, \ref{fig:uCOMP} and \ref{fig:Tcomp} show the results obtained in this manner for the specific volume decrease across the shock wave, shock wave speed and temperature for different initial compaction ratios $\eta_1$ and piston speeds $u_p$.  Also shown are the analytical predictions using the Helfand equation of state detailed in Section 2.  In all cases, the agreement was found to be excellent, with an error not greater than $\sim 3\%$, even for the large compaction ratio and strong shocks investigated.  Also noteworthy is the fact that the temperature jump across the shock wave displayed in Figure \ref{fig:Tcomp} shows the predicted invariance with the compaction ratio given by \eqref{Tinv}.

%%%%%%%%%%%%%%%%%%%%%%%%%%%%%%%%%%%%%%%%%%%%%%%%%%%%
\section{Conclusion}
The compressible dynamics and shock wave propagation in a dense hard disk medium, relevant to shock propagation in granular hydrodynamics or in liquids, was solved analytically using the Helfand equation of state.  Physical insight on the role of the medium's initial packing on the compressibility of the medium was greatly enhanced by having obtained closed form solutions for the isentropic exponent and shock jump conditions.  The analytical predictions were validated against molecular dynamics calculations using the Event Driven Molecular Dynamics technique, where the evolution of the system of colliding elements can be obtained analytically.  With increasing compaction of the medium, is was shown that the medium's compressibility changes substantially, with an isentropic exponent of $\gamma=2$ in the dilute gas phase, and $\gamma=\mathcal{O}(10)$ at higher compactions.  This was shown to affect significantly the shock Hugoniot and shock jump conditions.  Parametrization of the shock jump relations were obtained using the shock Mach number and piston speed.  The important result that the temperature in the compressed medium depends to a very good approximation only on the square of the piston speed for all compaction levels was shown analytically and demonstrated numerically via the molecular dynamic calculations. This important result provides a very simple means to estimate the amount of energy injected into a hard particle system by surfaces generating shocks. The present study assumed all collisions to be elastic; future study will be devoted to studying how the shock hydrodynamics are affected by the dissipative nature of granular flows or in reactive flows where collisions may be inelastic.

\section*{Acknowledgments}
We wish to acknowledge the financial support of the National Science and Engineering Research Council (NSERC) of Canada through a Discovery Grant to M.I.R., the support of the Defence Research and Development Canada - Suffield (Dr. Julian J. Lee as contract monitor), the partial support of the NSERC Hydrogen Canada (H2CAN) Strategic Research Network for supporting the undergraduate summer internship of N.S., the Work-Study program at the University of Ottawa for supporting one CO-OP session of N.S., and the University of Ottawa Initiation of Research grant to M.I.R.

\bibliographystyle{spphys}    
\bibliography{references}
\end{document}